# Hot Phonon and Carrier Relaxation in Si(100) Determined by Transient Extreme Ultraviolet Spectroscopy

**Scott K. Cushing[1,2], Michael Zürch[1], Peter M. Kraus[1], Lucas M. Carneiro[1,2], Angela Lee[1], Hung-Tzu Chang[1], Christopher J. Kaplan[1], Stephen R. Leone[1,2,3]**

[1]*Department of Chemistry, University of California, Berkeley, CA 94720, USA.*

[2]*Chemical Sciences Division, Lawrence Berkeley National Laboratory, Berkeley, CA 94720, USA.*

[3]*Department of Physics, University of California, Berkeley, CA 94720, USA.*

*Author e-mail address:* srl@berkeley.edu

**The thermalization of hot carriers and phonons gives direct insight into the scattering processes that mediate electrical and thermal transport. Obtaining the scattering rates for both hot carriers and phonons currently requires multiple measurements with incommensurate timescales. Here, transient extreme-ultraviolet (XUV) spectroscopy on the silicon 2p core level at 100 eV is used to measure hot carrier and phonon thermalization in Si(100) from tens of femtoseconds to 200 ps following photoexcitation of the indirect transition to the Δ valley at 800 nm. The ground state XUV spectrum is first theoretically predicted using a combination of a single plasmon pole model and the Bethe-Salpeter equation (BSE) with density functional theory (DFT). The excited state spectrum is predicted by incorporating the electronic effects of photo-induced state-filling, broadening, and band-gap renormalization into the ground state XUV spectrum. A time-dependent lattice deformation and expansion is also required to describe the excited state spectrum. The kinetics of these structural components match the kinetics of phonons excited from the electron-phonon and phonon-phonon scattering processes following photoexcitation. Separating the contributions of electronic and structural effects on the transient XUV spectra allows the carrier population, the population of phonons involved in inter- and intra-valley electron-phonon scattering, and the population of phonons involved in phonon-phonon scattering to be quantified as a function of delay time.**

## I. INTRODUCTION

Control of ultrafast carrier thermalization and transport processes is increasingly important in nanoscale semiconductor junctions [1], next-generation thermoelectrics [2], and hot carrier solar cells [3]. Through extensive optical and electrical characterization, the electron-phonon and phonon-phonon scattering processes have been detailed in Si, Ge, and GaAs for carriers in the lowest-lying conduction and valence valleys [4–6]. This information has proven vital for allowing accurate device prediction and modeling through the Boltzmann transport equations [7]. However, the time scale and energy range over which the individual scattering pathways can be tracked is limited by the narrow pump and probe pulse bandwidths required to select specific phonon or electron features. Additionally, the need for multiple types of instrumentation hinders rapid understanding of hot carrier transport and relaxation in nanostructured and two-dimensional materials.

Transient electron diffraction and x-ray diffraction measurements have made considerable progress towards understanding coupled carrier-phonon dynamics by directly measuring the lattice dynamics following photoexcitation [8–14]. The lattice deformations created by the initial carrier distribution and the lattice expansions caused by excitation of a non-thermal phonon bath during carrier thermalization have been measured and predicted in Si and other semiconductors at and above the melting threshold [15–21]. The electron-phonon scattering has been separated into three phonon modes using a non-thermal lattice model [8]. Coherent optical and acoustic phonon measurements, as well as acoustic shock wave measurements, have also allowed phonon creation and decay to be understood following electronic excitation [22–28]. The lattice dynamics are directly measured in each of these investigations, but the electronic contribution must often be inferred. A table-top technique that can directly measure both the carrier and phonon distributions following excitation remains to be established.

Generation of extreme ultraviolet light (XUV) by high harmonic generation (HHG) can be used to probe the electronic and structural dynamics through core-level transitions, similar to previous measurements at synchrotron and free electron sources but using a table-top setup [29–34]. When a core electron is promoted to an unoccupied state, the core-hole potential modifies the valence potential, and a highly localized core-hole exciton is formed. The measured XUV absorption is therefore distorted from the ground-state unoccupied density of states and contains local structural information [35–37]. In atomic and molecular systems, core-hole effects can be theoretically predicted, allowing electronic and vibrational dynamics to be quantified following photoexcitation [38,39]. In a semiconductor, multi-electron and many-body effects complicate the interpretation and prediction of XUV absorption, making it difficult to separate electronic and structural contributions [40–42]. This has so far slowed the use of transient XUV spectroscopy as a single-instrument method for understanding carrier and phonon thermalization pathways in semiconductors.

In this article, the underlying electronic and structural contributions to the Si $L_{23}$ edge evolution are separated following 800 nm optical excitation to the Δ valley. Ground and excited state calculations using a single plasmon pole model and the Bethe-Salpeter equation (BSE) with density functional theory (DFT) are used to interpret the measured changes in the Si $L_{23}$ edge XUV absorption. Hot carrier thermalization dynamics are resolved through state-filling at the appropriate valleys' critical points. Lattice dynamics are recognized using the unique changes in the critical point structure that result from optical and acoustic phonon excitation. From the comprehensive measurements of XUV transient absorption versus time, a [100] lattice deformation is extracted with kinetics that mirror the high-energy phonons involved in inter-valley electron-phonon scattering. Additionally, a thermal lattice expansion is obtained with kinetics that follow the creation of low-energy, mainly acoustic phonons by intra-valley electron-phonon scattering and phonon-phonon decay processes. These findings suggest that ultrafast pump-probe transient XUV spectroscopy can provide the important carrier and phonon scattering timescales and pathways following photoexcitation in a single set of measurements.

## II. METHODS

### A. Experimental

The static and transient XUV absorption spectra of 200 nm thick, p-type (B-doped, $10^{15}$/cm$^3$) Si (100) membranes purchased from Norcada are measured with high-harmonic generation (HHG) XUV radiation. The HHG is produced in helium gas (semi-infinite gas cell) with a 50 fs pulse duration, 1 kHz repetition rate regeneratively amplified Ti-sapphire laser [32]. A 400 nm frequency doubled output produced in a BBO crystal is added to the 800 nm fundamental to produce both even and odd harmonics in the HHG process. The excess 800 nm and 400 nm light is removed from the HHG spectrum using a combination of a micro-channel plate (MCP) and Zr filters, resulting in XUV harmonics spanning from 70-120 eV [43]. The excitation wavelength is chosen to match the indirect optical transitions to the Δ (800 nm, 1.55 eV) valley [44]. The polarizations of the pump and probe are parallel to the [110] direction of the (100) membrane. Delay times between optical pump and XUV probe are obtained by varying the distance of a retroreflector with a computer-controlled delay stage.

Pump powers are adjusted to produce a carrier density of 1.5 x $10^{20}$/cm$^3$. The average carrier density ($\Delta N$) is estimated using [13]

$$\Delta N = \frac{F}{\hbar\omega}\frac{1-R}{L}(1-\exp(-\alpha L))(1+R\exp(-\alpha L)), \qquad (1)$$

where $F$ is the laser fluence, $R$ is the reflectivity of the thin film, $\hbar\omega$ is the energy of the photons, $L$ is the membrane thickness, and $\alpha$ is the absorption coefficient for Si [45]. The second exponential term accounts for back-reflections at the rear membrane-vacuum interface. The absorption was estimated as 2.5±0.5x$10^3$ cm$^{-1}$ using the measured transmission of the sample at 800 nm, and then accounting for reflectivity losses using the Fresnel equations for thin films and the known refractive index of silicon. To prevent the possible propagation of error from this estimated absorption, the excited carrier density is fit during the analysis. For 800 nm excitation, the absorption depth is larger than the 200 nm membrane thickness, so the depth-dependent effects from the photoexcited carrier distribution are negligible.

The effects of two-photon absorption (TPA) are considered for the high fluence used, especially for 800 nm excitation across the indirect band gap [46,47]. The spatially- and intensity-dependent absorption can be solved for using

$$\frac{dI}{dz} = -\alpha I - \beta I^2, \qquad (2)$$

where $I$ is the peak intensity of the sech$^2$ experimental pulse, $\alpha$ is the same as in Equation (1), and $\beta$ is taken as 2x$10^{-9}$ cm/W [48,49]. Given that one electron-hole pair is created for every two-photons in two-photon absorption, Equation (2) predicts that at 800 nm excitation the two photon absorption contribution is less than 10% of the photoexcited carriers at the intensities used in the experiments. The lack of a two-photon absorption contribution is also experimentally confirmed by the absence of a measurable state-filling signal within the 5 mOD noise level at the $L_1$ critical point of the XUV spectrum for 800 nm pulsed excitation. Two photon absorption effects are therefore excluded in the experimental analysis, recognizing that the <10% modulation would be below the 5 mOD noise level of the <40 mOD overall changes measured in the experiment.

### B. Theoretical

Calculation of the ground state XUV absorption is done within the OCEAN code (Obtaining Core-level Excitations using Ab initio methods and the NIST BSE solver) [50,51]. The density functional level is used to calculate the ground state electron densities and wave-functions using Quantum-ESPRESSO [52]. In the OCEAN code, projector augmented wave (PAW) reconstructed wave functions are used to estimate the core-level transition matrix elements. The dielectric screening is estimated using a real-space random phase approximation inside a sphere around the atom along with a Levine-Louie model dielectric function outside the sphere [53,54]. The final electron-hole states are then calculated by the Bethe-Salpeter (BSE) equation.

The specific parameters for the DFT and BSE-DFT calculations are as follows. The DFT density of states is calculated within the local density approximation (LDA) using a norm-conserving generalized gradient approximation (GGA) Perdew-Burke-Ernzerhof pseudopotential with a converged k-point mesh of 20x20x20 points and a plane wave cutoff of 100 Ry. The converged lattice constant was 5.46 Angstroms. The BSE-DFT calculations in OCEAN are then performed using the Quantum ESPRESSO results. The final and projector augmented wave states are found converged to an accuracy that reproduces the experimental absorption at k-point meshes of 8x8x8 and 2x2x2, respectively. The total number of bands for the final and projector augmented wave states are well converged at 100 and 200, respectively. The SCF mixing is taken as 0.7 with 250 iterations used. The BSE mesh is taken as 6x6x6 and a cut-off radius of 8.0 Bohr is used. A radius of 8.0 Bohr is also used for the projector augmented wave shell with a 0.8 scaling factor of the slater G parameter. A dielectric constant of 11.7 is used for silicon. The absorption is calculated for XUV dipole orientations along the [100] and [110] directions. Within the experimental broadening, little difference in the predicted spectrum is found between these dipole orientations.

The varying lattice expansions for the non-linear fit are linearly interpolated from a series of lattice expansions in the BSE-DFT calculations to make the fitting procedure computationally feasible. Lattice expansion values of -5% to 5% are computed to allow an accurate interpolation of intermediate values and to not restrict the non-linear fit parameter search space. The unit cell is relaxed for each expansion value. The BSE-DFT calculations do not properly describe valence core-hole effects at energies below the Fermi level or interference from photoelectron scattering pathways that begin at 5-10 eV above the Fermi level. The best fit is therefore determined in the energy range of 98-105 eV. The loss in accuracy outside this energy range is reflected in the discrepancies between the experimental and predicted ground state absorption shown in Figure 1b. The inaccurate description of the valence region, the weak Si *2p* transition strength to the mostly *p*-character valence band, and the excited-state broadening prevent the hole signatures from being extracted with reasonable certainty.

## III. ELECTRONIC AND STRUCTURAL CONTRIBUTIONS TO THE XUV ABSORPTION

### A. Ground State Absorption

An accurate mapping of the silicon band structure onto the ground state XUV absorption is critical for interpreting the experimental spectra. Not only does the photoexcited core-hole modify the energy and amplitude of the band structure's critical points, but the energy-dependent core-hole lifetime smooths the critical point structure at energies higher than the transition edge. By first modeling the effects of the Si *2p* core-hole on the ground state spectrum, the measured excited state spectrum can be more accurately interpreted in terms of the photoexcited carrier and phonon distributions. The amplitudes and energies of the $L_1$ and $\Delta_1$ critical points are otherwise incorrectly assigned in the ground state XUV absorption; a discrepancy which would prevent an accurate interpretation of the excited state XUV spectrum.

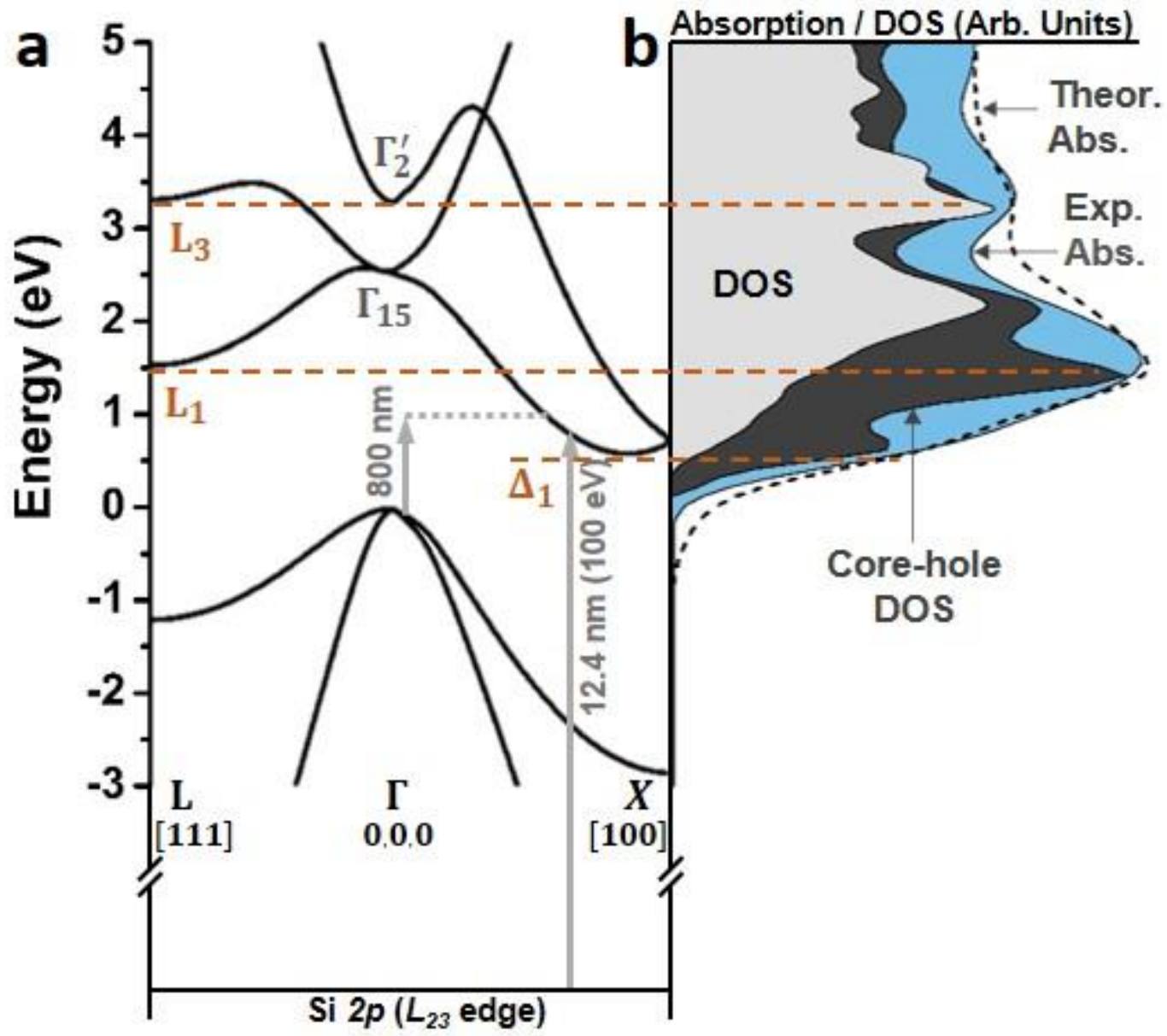

***Figure 1***. *Critical points and core-hole modification of the spectrum of the Si 2p $L_{23}$ edge. (a) Band structure of silicon along the L-Γ-X path, highlighting the $\Delta_1$, $L_1$, and $L_3$ critical points. The k-space directions are marked. Note L is at ½,½,½ and Δ is at ~0.8,0,0 in the Brillouin zone. The excitation wavelength and an example core level transition are marked as arrows. The arrow length is adjusted for the underestimation of the band gap in the DFT calculation. (b) Comparison of the* s+d *projected density of states (DOS, grey), core-hole modified DOS (dark grey), experimental absorption (blue), and theoretical absorption (dashed) predicted by the core-hole modified DOS broadened with a single plasmon pole model. The 2p core-hole exciton renormalizes the $L_1$ and $\Delta_1$* s-p *hybridized critical points, while having less effect on the higher lying, d-character $L_3$ critical point.*

To facilitate comparison between the measured ground state XUV spectrum and the band structure of silicon, the measured XUV absorption spectrum is deconvoluted for spin-orbit splitting in Figure 1 using two delta functions spaced by 0.6 eV [36]. The DFT predicted critical points of the band structure along the L-Γ-X path and the *s+d* projected density of states (DOS) relevant to the Si *2p* $L_{23}$ transition are compared to the measured XUV absorption in Figure 1a and 1b. The oscillator strength for the *s* character states is assumed to be twice that of the *d* character states [36]. Figure 1b shows that the experimental ground state XUV absorption does not directly map onto the ground-state DOS calculated by DFT. In particular, the amplitude of the absorption near the $L_1$ and $\Delta_1$ critical points is underestimated. This discrepancy occurs because the *2p* core-hole excited by the XUV transition alters the valence potential, creating a core-hole exciton and modifying the DOS, especially for the *s-p* hybridized $L_1$ and $\Delta_1$ critical points [36,37].

The effects of the core-hole on the final DOS in the XUV transition can be calculated using the Bethe-Salpeter equation (dark grey area in Figure 1b) [50,51]. The BSE-DFT calculation accurately predicts that the Si *2p* core-hole distorts the projected *s+d* DOS, especially at the $\Delta_1$ and $L_1$ points. The BSE-DFT calculation also uses projector augmented wave (PAW) dipole transition elements instead of the estimated *s+d* contribution of the DFT only calculation. Following the BSE-DFT calculation, the theoretical XUV absorption is calculated by broadening the core-hole modified DOS (dashed line in Figure 1b). The XUV broadening is treated as an energy-independent core-hole lifetime plus an energy-dependent lifetime that depends on the excited core-level electron's inelastic scattering [50,51,55–57]. The inelastic electron scattering can be modeled using an empirical model, such as the Seah-Dench formalism, or by using the imaginary part of the self-energy to calculate the electron's inelastic mean free path [58–61]. Here, the latter approach is taken to allow the effects of visible light photoexcitation on the core-level transition to be included when modeling the transient absorption data sets.

Specifically, the energy-dependent broadening is approximated using a Drude-Lindhard single-plasmon pole model, which can accurately represent the experimental Si electron energy loss function [62,63]. A Gaussian instrumental broadening of 0.2 eV and an energy-independent core-hole lifetime represented by an energy width of 15 meV are also included [64]. Silicon's valence electron density sets the plasmon pole in the electron energy loss function at ~16.8 eV above the Fermi level. The single plasmon pole model of the electron energy loss function is parameterized as [59]

$$Im\left[-\frac{1}{\epsilon(\Delta E,k)}\right] = \frac{A_1\gamma_1\Delta E}{\left(\left(\hbar\omega_{0,k}\right)^2-\Delta E^2\right)^2+\gamma_1^2\Delta E^2} * \Theta(\Delta E - Eg)\ , \qquad (3)$$

with

$$\hbar\omega_{0,k} = \hbar\omega_0 + C * \frac{\hbar^2k^2}{2m}, \qquad (4)$$

where $\Delta E$ is the energy difference between the Fermi level and the core-level excited electron, $k$ is the momentum of the electron, $Eg$ is the band gap, $A_1$ is the oscillator strength equal to 256.4 eV$^2$, $\gamma_1$ is the damping coefficient equal to 3.8 eV, $\hbar\omega_0$ is the energy of the plasmon pole equal to 16.8 eV, $m$ is the electron mass, and $C$, here taken as 0.5, is a factor used to approximate the parabolicity of the conduction bands. The core-level transition lifetime is then calculated in terms of the inelastic mean free path, $\lambda$, of the electron as [59]

$$\lambda^{-1}(E) = \frac{1}{\pi E}\int dE' \int dk \frac{1}{k} Im\left[-\frac{1}{\epsilon(E\prime,k)}\right], \qquad (5)$$

where the energy integral goes from 0 to $E - E_f$ and the momentum integral goes from $k_\pm = \sqrt{\frac{2mE}{\hbar^2}} \pm \sqrt{\frac{2m}{\hbar^2}(E-E')}$. These bounds limit the scattering space possible for the core-level excited electron. The inelastic mean free path is converted into an energy dependent lifetime $\Gamma(E)$ using

$$\Gamma(E) = \frac{\hbar}{\lambda(E)}\sqrt{\frac{2E}{m}} + \Gamma_{CH}\ , \qquad (6)$$

where $\Gamma_{CH}$ is the 15 meV core hole lifetime [64]. The XUV absorption is then calculated by convolution of a Lorentzian with energy-dependent width given by Equation (6) and the core-hole modified DOS calculated in Figure 1b.

Equation (5) represents the increasing loss channels of the photoexcited core-level electron with increasing energy. The absorption spectrum becomes broader with increasing energy because more loss channels are possible. The single plasmon pole model is an approximation, however, which neglects the more complicated valence loss channels near the conduction band edge. Further, the Drude-Lindhard approximation is only valid at small energy and momentum ranges above the Fermi level. These approximations lead to an over-estimation of the broadening a few eV above the Fermi level, which creates an under-estimation in depth for the above-edge minima in Figure 1b. The single plasmon pole model is used because it allows the change in the energy-dependent broadening to be estimated following visible light photoexcitation from the photoexcited carrier density. If a non-energy-dependent broadening is used, the fine features near the XUV edge and broad features above the XUV edge cannot be predicted in the ground state spectrum, nor can the excited state broadening change be predicted from the photoexcited carrier density. The calculated absorption in Figure 1b accurately predicts the measured static Si $L_{23}$ absorption within these approximations.

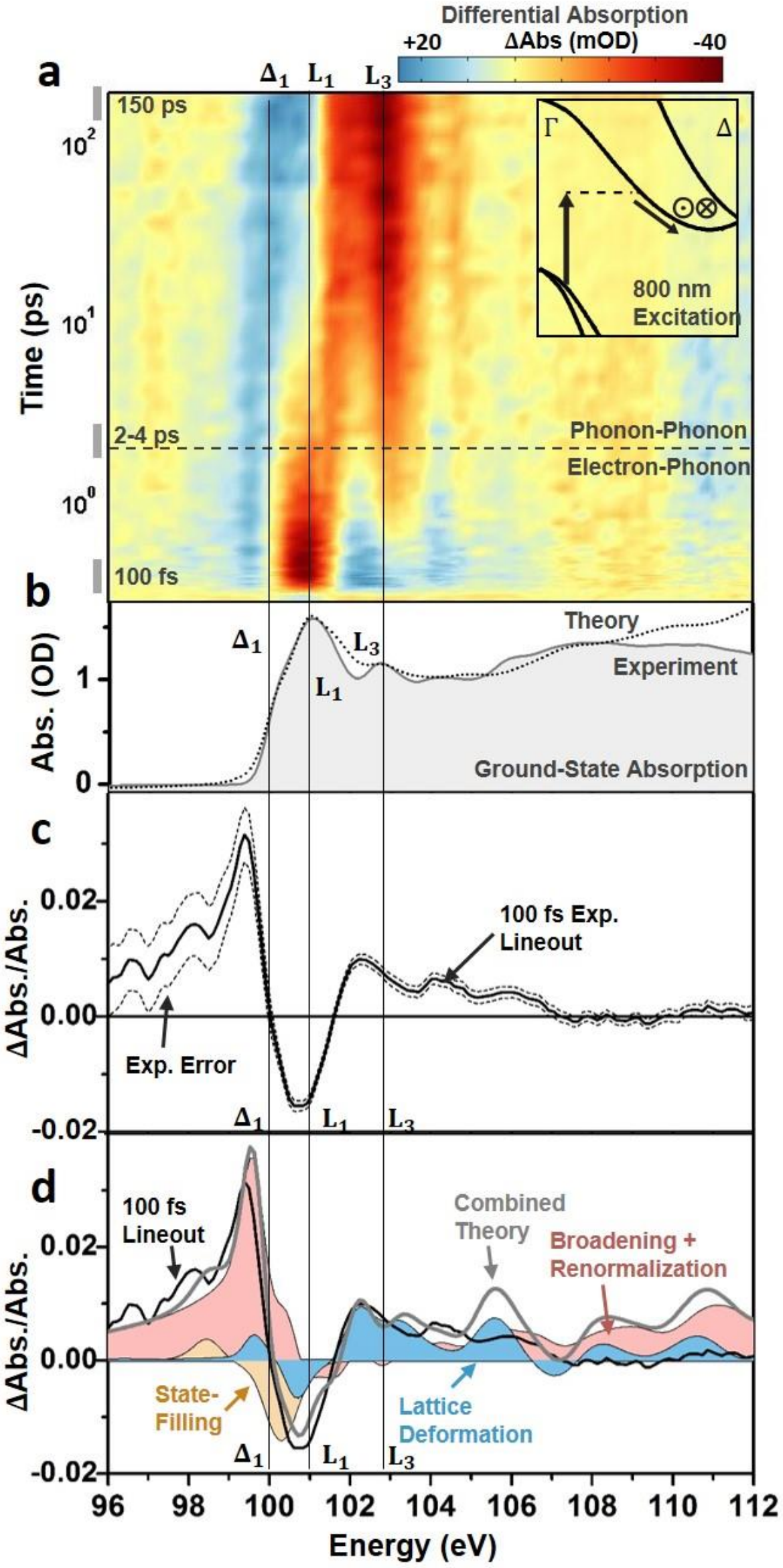

***Figure 2***. *Differential absorption data following 800 nm excitation of the Si* $L_{23}$ *XUV edge. (a) The differential absorption data are shown as a color map on a logarithmic time scale up to 200 ps time delay following 800 nm excitation to the* $\Delta_1$ *valley. The inset shows the excitation and scattering pathways for the excited electrons. The in and out of plane arrows indicate where inter-valley scattering between degenerate valleys is possible. The cross-over time between predominantly electron-phonon scattering or phonon-phonon scattering is also indicated by the dashed horizontal line. (b) The static ground state experimental and theoretical absorption from Figure 1b are shown for comparison to the differential absorption. (c) The experimental 100 fs differential absorption (black solid line). The dashed lines indicate the 95% confidence intervals of the measurement. (d) The theoretically predicted state-filling (light orange), broadening and renormalization (light pink), and lattice deformation (light blue) contributions for a* $1.5 \times 10^{20}/cm^3$ *photoexcited carrier density The grey solid line indicates the combined theoretical contributions of the shaded areas.*

## B. Excited State Absorption for 800 nm Excitation

An experimental differential absorption spectrum of the Si $L_{23}$ edge absorption versus time following photoexcitation with 800 nm light is shown in Figure 2a. The differential absorption is reported as the logarithm of the pump-on divided by the pump-off spectrum, and the major features of the spectrum align with the critical points of the ground state XUV absorption (Figure 2b). Similar to modeling the ground state absorption, the core-hole effects and energy-dependent broadening must be considered when interpreting the measured excited state spectrum. In the excited state spectrum, the change in carrier density following visible light photoexcitation results in both electronic and structural changes to the silicon (Figure 2c). The electronic effects can be categorized as state-blocking, broadening, and renormalization, all of which can be considered instantaneous within the 50 fs 800 nm pulse width. The structural changes can be categorized as lattice deformations and expansions resulting from carrier-phonon and phonon-phonon scattering. The different electronic and structural contributions are summarized in Figure 2c for the 100 fs time slice and detailed in the following sub-sections.

### *1. Electronic Contributions to the Excited State Absorption*

As noted, the electronic contributions can be summarized as photoinduced state-filling, broadening, and renormalization. First, the promotion of carriers from the valence to conduction band leads to changes in occupation of the band structure of silicon, modulating the absorption spectrum [4,65]. Here, this effect is referred to as state-filling, which describes the allowing or blocking of XUV transitions from the Si *2p* core level to the valence or conduction band, respectively, following the promotion of electrons by the 800 nm pump pulse (Figure 2c). For the Si $L_{23}$ edge, the transition edge lies at approximately 100 eV [66]. Specifically, the $L_2$ onset energy is taken as 99.8 eV following the spin-orbit deconvolution. Immediately following the excitation pulse, holes created in the valence band are expected to increase the XUV transition rate at energies lower than 98.7 eV, or the edge onset energy minus the band gap energy. Similarly, the optical promotion of electrons to the critical points in the conduction band is expected to decrease the absorption above 100 eV. Specifically, for the 800 nm excitation into the $\Delta_1$ point, a decrease in absorption is expected at 100.25 eV, which is the valence band maximum at 98.7 eV plus the 1.55 eV pump energy.

The possible state-filling contributions are modeled to first order by first blocking (allowing) transitions in the core-hole modified DOS before broadening. The 0.2 eV experimental resolution is close to the 0.3 eV width of the excitation spectrum of the 800 nm pump, obscuring fine detail in the state-blocking spectrum. This allows a Gaussian distribution of carriers with 0.3 eV width to be used for both the initial non-thermal and final thermalized carrier distribution. The percentage of state-filling is normalized by the calculated DOS to account for Pauli-blocking near the CB and VB edge [67]. The XUV blocking percentage will depend on the orbital character of the band. Since the XUV transition occurs from the *p*-character core level, unequal differential absorption magnitudes exist for electrons in the mixed *s-p* character conduction band and holes in the mainly *p*-character valence band. The calculated state-filling percentage is also adjusted for the character of the probed DOS in the dipole-allowed XUV transitions.

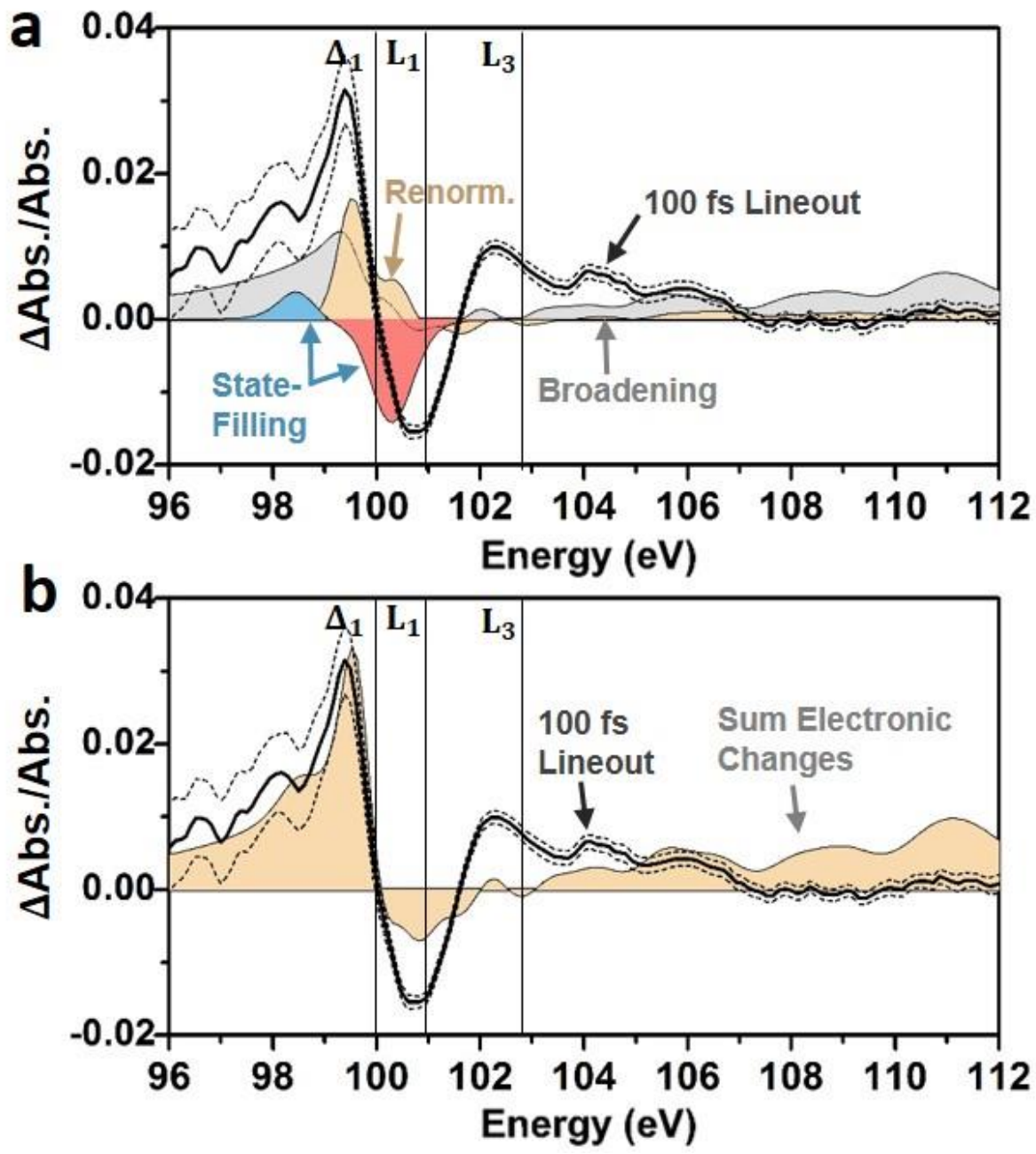

***Figure 3**. Electronic contributions to the excited state XUV absorption. (a) The theoretically predicted state-filling of holes (light blue) and electrons (light pink), broadening (grey), and renormalization (light orange) contributions for a 1.5 x $10^{20}$/cm$^3$ photoexcited carrier density. The theoretical predictions are compared to the experimental 100 fs differential absorption (black solid line). The dashed lines indicate the 95% confidence intervals of the measurement. (b) The orange shaded area indicates the sum of the electronic contributions predicted in panel (a). The broadening, renormalization, and state-filling combine to accurately predict the increased absorption at energies below the $\Delta_1$ point. The decreased absorption at the $L_1$ point is predicted at the correct energy, but the amplitude of the $L_1$ and the $L_3$ features are underestimated when only the electronic contributions are considered.*

Using the ground-state absorption model of Figure 1 and a carrier density of 1.5x$10^{20}$ carriers/cm$^3$, the photoexcited carriers are predicted to create an increased absorption below 100 eV (light blue region in Fig. 3a) and a decreased absorption above the $\Delta_1$ critical point that is centered around 101 eV (light pink region in Fig. 3a). The decreased absorption predicted in Figure 3a does not energetically align with the experimentally measured decrease in absorption (black solid line in Figure 3a). This is because the excited state broadening, renormalization, and structural effects discussed in the following paragraphs partially cancel the change in absorption from the state-filling contribution (Figure 3b). It should be noted that in the first 50 fs electron-electron scattering can energetically broaden but not relax the excited carrier distribution. The impact of this energetic redistribution on state-filling is not discernable within the 50 fs pulse-width and excited state and instrumental broadening.

In addition to state-filling effects, the reduction in valence electron density and increase in the conduction electron density will also change the excited state XUV broadening. The change in the ground state XUV broadening with photoexcitation is modeled to first order by changing the carrier density in Equation (3). Specifically, the experimentally excited carrier density of 1.5x$10^{20}$ carriers/cm$^3$ decreases the plasmon pole frequency by 0.16 eV through

$$\omega_0 = \sqrt{\frac{(n-\Delta N)q^2}{m\epsilon_0}}, \qquad (7)$$

where $n$ is the ground state valence carrier density, $m$ is the mass of an electron, $\epsilon_0$ is the permittivity of free space, $q$ is the charge of an electron, and $\Delta N$ is the excited carrier density. Decreasing the valence plasmon by the photoexcited carrier density from 16.8 eV to 16.64 eV increases the energy-dependent broadening following integration in Equation (5). In the differential absorption spectrum, the increase in broadening (grey area in Figure 3a) leads to an increase in absorption at energies above and below the $L_1$ critical point. As noted in Section III-A, treating the broadening by this approach ignores the 800 nm photoexcited conduction band plasmon as well as modification of other valence loss channels.

Finally, changing the valence and conduction electron densities will change the screening of the electrons, creating a photoexcited renormalization of the band gap [68–71]. The magnitude of the band gap renormalization is over-estimated using only a plasmon-pole model [72], so here the predicted cube root functional dependence of the carrier density is scaled to the experimentally reported values [73]

$$\Delta E_{CB}(\Delta N) = -D * \left(\frac{\Delta \mathrm{N}}{\mathrm{N}_0}\right)^{\frac{1}{3}}, \qquad (8)$$

which gives a downward shift of the conduction bands ($\Delta E_{CB}$) of ~60 meV at 1.5 x $10^{20}$/$cm^3$. The valence band shifts by a similar amount, but valence band effects are not clearly observed in the experimental spectra. In Equation (8) $\Delta N$ is the excess carrier density while D and $N_0$ are fit parameters of values 0.05 eV and 1x$10^{20}$ $cm^{-3}$, respectively, for the carrier density range used in these experiments.

To completely describe the renormalization of the XUV absorption edge, the change in screening of the core-hole exciton must also be estimated to first order following Reference [74] by

$$\Delta E_{core-hole}(\Delta N) = \frac{3q^2}{(4\pi)^2 \epsilon \epsilon_0}(3\pi^2 \Delta N)^{\frac{1}{3}}, \qquad (9)$$

where $\Delta N$ is the excess carrier density, $\epsilon$ the dielectric constant, $\epsilon_0$ the vacuum permittivity, and $q$ the charge of an electron. The renormalization of the band gap and core level transition are then approximated by uniformly shifting the core-hole adjusted DOS in the ground-state model by

$$\Delta E(\Delta N) = \Delta E_{core-hole}(\Delta N) + \Delta E_{CB}(\Delta N). \qquad (10)$$

The decrease in binding energy of the core-hole exciton (Equation (9)) counteracts the band gap renormalization (Equation (8)) to large extent, leading to a smaller overall renormalization of the XUV transition of <10 meV instead of 60 meV. The smaller renormalization of the Si $L_{23}$ edge compared to the band gap has been experimentally confirmed by doping studies [75]. The uniform shift of the conduction band structure is an approximation based on Reference [76] where a many-body perturbative GW calculation was used to predict the excited state XUV spectrum to similar overall effect. In both cases, the renormalization primarily leads to an increase in absorption around the $\Delta_1$ point (orange area in Figure 3a).

It should be emphasized that Equations (7) - (10) are based on first order approximations of the many-body response to the photoexcited electron density as verified by visible light pump-probe experiments, ignoring the possibility of a core-hole perturbation to the many body response. The x-ray probe excitation density is one million times less than the visible light pump, making this a plausible assumption. The validity of this assumption can be evaluated by comparison to other x-ray absorption measurements. In particular, identical to visible light experiments, attosecond x-ray absorption experiments confirm that the electronic response is complete within <20 fs [76]. During this time, the change in the differential absorption can be completely explained by electronic contributions. The structural response from photoexcited electrons was measured to only be significant on >30 fs timescales, at which time the spectral features match those in Figure 2a. The spectral features of this short time scale strain are further confirmed by comparison to steady-state x-ray absorption measurements of strained silicon membranes [77]. Similarly, the spectral features of the long time scale isotropic lattice expansion is confirmed by comparison to heated silicon [78]. While these experiments seem to confirm the use of these first order approximations, it still must be emphasized that these calculations are approximations, and the full many-body response of the valence to the photoexcited electrons and core hole needs to be further explored.

Overall, in the predicted XUV differential absorption, the electronic contributions sum to create a below-edge increase and an above-edge decrease in the XUV absorption (Figure 3b). The electronic components also contribute to the broad increase in absorption at energies greater than 103 eV. The combined renormalization and broadening contributions span the entire measured energy range, whereas state-filling contributions occur mainly near the transition edge. At the transition edge, the electronic contributions have similar amplitude but different signs, leading to cancellation. The cancellation between the different electronic contributions in Figure 3 emphasizes the importance of modeling how the ground state band structure relates to the XUV absorption. Otherwise, the region of decreased absorption below 100 eV and increased absorption between 100 eV and 102 eV could incorrectly be assigned as only state-filling from electrons and holes. The remaining discrepancy between the measured and predicted differential absorptions at energies above 100 eV suggests the presence of structural contributions, as discussed in the next section.

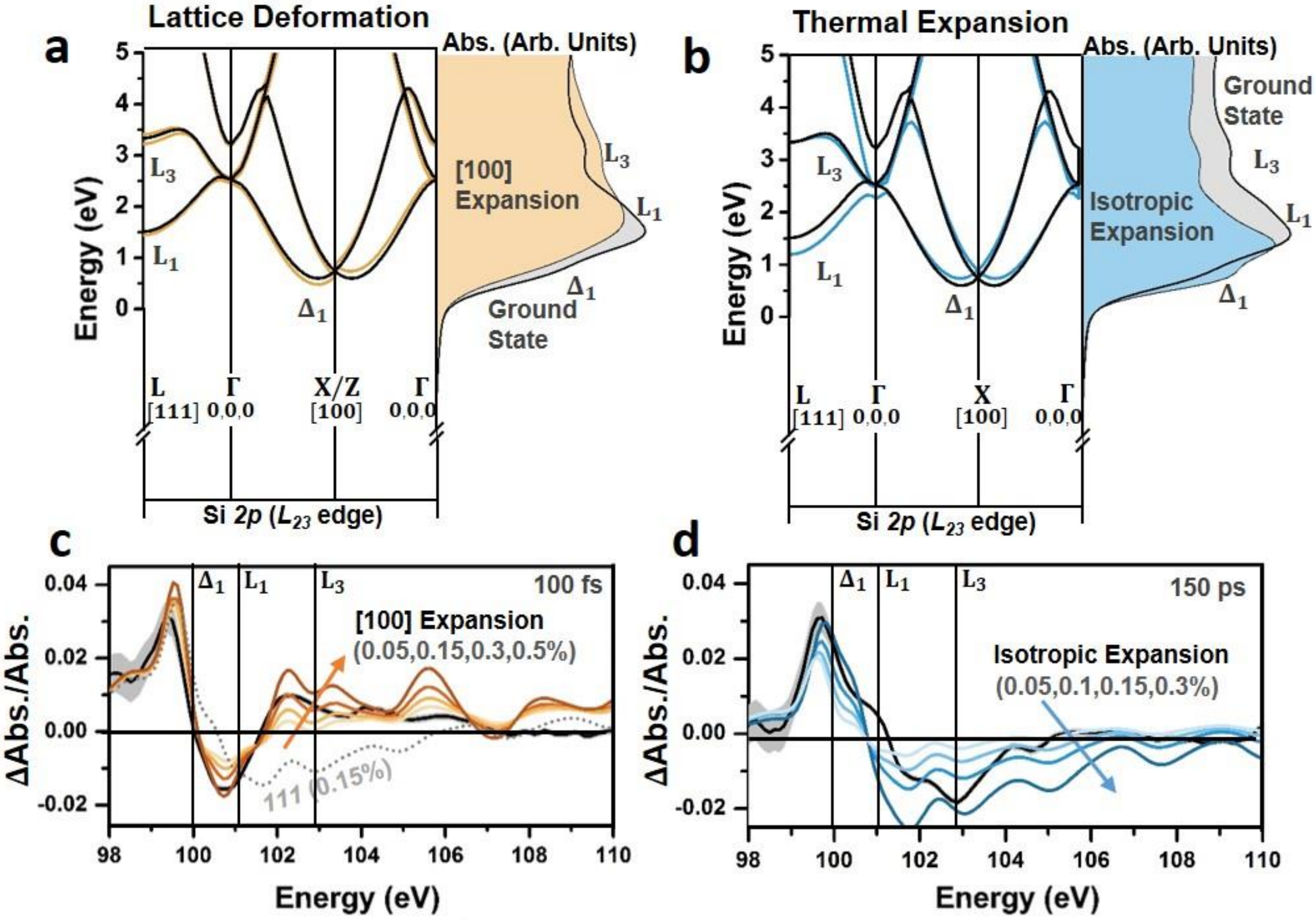

***Figure 4.*** *Structural contributions to the excited state XUV absorption. (a) The predicted effect of an expansion along the [100] direction (light orange) and (b) an isotropic lattice expansion (light blue) on the Si band structure, shown as the resulting change in the theoretically predicted absorption. In each case a 3% expansion of the relevant axis is used. The k-space directions are marked. Note L is at ½,½,½ and the Δ is at ~0.8,0,0 in the Brillouin zone. The ground state silicon band structure (black line) and ground state absorption (grey area) are shown for reference to the [100] and isotropic expansion calculations (colored lines and colored areas). The top of the valence band is arbitrarily set to zero. Comparison of the experimental (solid black line) and theoretical (color-range or dashed line) differential absorption lineouts at (c) 100 fs and (d) 150 ps after 800 nm excitation. The theoretical differential absorption is calculated using a range of [100] expansions for the 100 fs time slice and a range of isotropic expansions for the 150 ps time slice. In each case the carrier densities from Table I are used. An isotropic expansion is also shown for comparison for the 100 fs time slice as a grey dotted line. The grey shaded areas show the 95% confidence intervals on the experimental data*

### *2. Structural Contributions to the Excited State Absorption*

The first-order approximations of the electronic contributions following visible light photoexcitation only partially predict the measured changes in the ground state absorption (Figure 3b). The differential absorption above 100 eV is not replicated in the modeled spectrum. The above edge spectral features primarily relate to the local structural environment [35,37]. The incomplete modeling of higher energy features suggests that the dynamics of the local structural environment may also need to be considered. Excited state structural changes are possible by the anharmonicity of excited optical and acoustic phonon modes, as well as the screening of bonds by the photoexcited carrier density, creating lattice deformations and expansions. Depending on the axis of the lattice deformation or expansion, distinct changes will occur to the critical points of the Si $L_{23}$ edge as shown in Figure 4 [77,79–81].

For example, a [100] lattice deformation redshifts the Δ, Γ, and L valleys with a magnitude dependent on their location in the Brillouin zone relative to the direction of the applied strain (Figure 4a). The [100] lattice deformation also splits the six degenerate $\Delta_1$ valleys by approximately 150 meV for each 1% of expansion [77,79–81]. Together, these two effects result in a decreased (increased) absorption at energies below (above) the $L_1$ critical point in Figure 4a. Unlike the [100] lattice deformation, an isotropic expansion does not break degeneracy or introduce different relative shifts to the critical points in the XUV spectrum (Figure 4b). This results in a more uniform spectral redshift of the Si $L_{23}$ edge, leading to an increased (decreased) absorption at energies below (above) the $L_1$ critical point in Figure 4a [78].

The electronic effects of Section III-1 are combined with a range of [100] lattice deformations and isotropic lattice expansions in Figure 4c and 4d, respectively. An isotropic lattice expansion is distinct from the differential absorption features of an anisotropic lattice deformation, as can be visualized by comparing the colored lines to the grey dashed line in the Figure 4c. The combined electronic and structural contributions seem to improve agreement between the modeled and measured differential absorption at both 100 fs and 150 ps. However, the time-dependent magnitude of the possible structural contributions will depend on the evolution of the initially excited carrier distribution. This means that, unlike the electronic effects discussed in Section IIIB-1 which were complete within the pulse width, the magnitude of the structural effects cannot be easily estimated from the initial experimental excitation alone. A best-fit procedure is instead applied to determine the presence of any structural contributions at three time scales following photoexcitation (Figure 5). This approach assumes that an equilibrium lattice deformation can approximate any non-equilibrium effects. The excited carrier density, lattice deformation, and lattice expansion are assessed by a non-linear fit using a robust algorithm weighted by the experimental uncertainty. Other combinations of biaxial or triaxial strains could not fit the experimental differential absorption. The fit results are summarized in Table I.

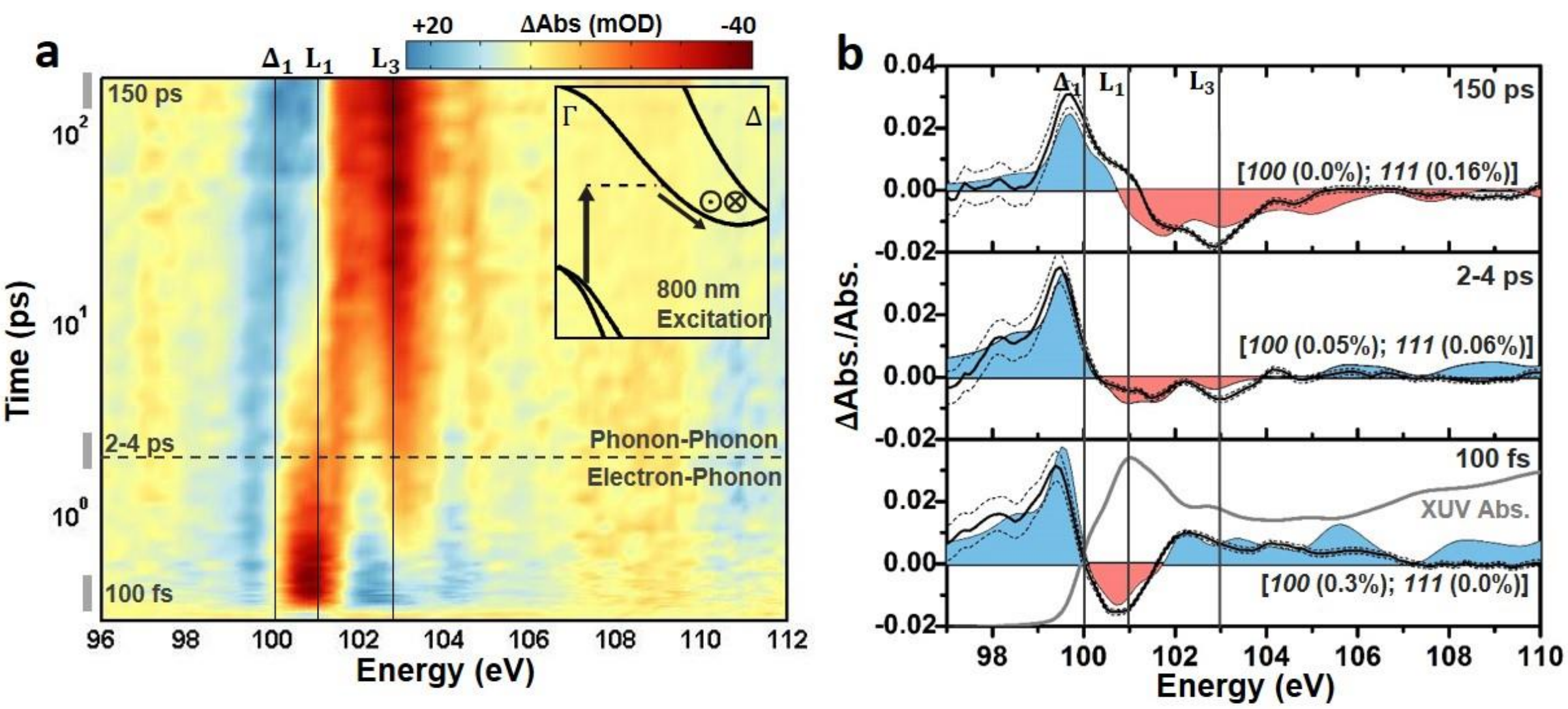

***Figure 5.*** *Differential absorption of the Si* $L_{23}$ *edge and comparison to theory. The differential absorption is shown on a logarithmic time scale from 0 to 200 ps for 800 nm excitation to the* $\Delta_1$ *valley. The inset represents some of the possible excitation and scattering pathways for the excited electrons. The in and out of plane arrows indicate where inter-valley scattering between degenerate valleys is possible. The cross-over time between predominantly electron-phonon scattering or phonon-phonon scattering is also indicated by the dashed horizontal line. A time slice (times indicated by gray bars to the left of panel a) from these periods is shown in (b). (b) The theoretical predictions from the single plasmon pole and BSE-DFT calculation are shown as the red and blue colored shading. The color shading matches the increased and decreased absorption in (a). The best fit percentage expansion of the relevant lattice vectors is indicated. The best fit range covers 98-105 eV, above and below which the BSE-DFT calculation loses accuracy. In (b) the ΔAbs./Abs. scale is used to allow direct comparison of experiment (solid line) to theory (shaded area) without scaling of the results. The dashed lines indicate the 95% confidence intervals of the measurement. The log scale of time in (a) is offset by 100 fs for visualization. The lineouts in (b) are averaged over the four nearest time-points.*

**Table I** *Non-linear Fit Carrier Densities and Lattice Expansions. Fit quantities are for an average of four times around the time indicated. 100 expansion denotes an expansion along the [100] direction while a 111 expansion denotes an isotropic expansion along each axis.*

| | | **~100 fs** | **~2-4 ps** | **~150 ps** |
|---|---|---|---|---|
| **800 nm (Δ)** | **Carrier Density (x10$^{20}$ /cm$^3$)** | 1.5±0.1 | 1.5±0.1 | 0.6±0.1 |
| | ***100*** **Expansion (%)** | 0.3±0.05 | 0.05±0.03 | 0.00±0.08 |
| | ***111*** **Expansion (%)** | 0.0±0.01 | 0.06±0.01 | 0.16±0.02 |

***(a) 100 Femtoseconds*** In the first several hundred femtoseconds following 800 nm excitation, the electrons redistribute between the degenerate Δ valleys by inter-valley scattering with optical phonon modes [82–85]. As shown in the bottom panel of Figure 5b, a 100 fs differential absorption time slice is best fit by a [100] expansion of 0.3±0.05%, no isotropic expansion at 0.0±0.01%, and a carrier density of 1.5±0.1 x10$^{20}$ carriers/cm$^3$. When combined with the electronic contributions, the [100] deformation accurately replicates the increased absorption at energies below the $\Delta_1$ critical point, the decreased absorption at the $L_1$ critical point, and the increase in absorption above 102 eV. Discrepancies between the theoretical and experimental differential absorption are greatest at the critical points that are not accurately described in the ground state spectrum.

For 800 nm excitation polarized along the [110] axis, dipole selection rules for the indirect transition lead to the four parallel and two perpendicular Δ valleys being non-equally populated [86–89]. A similar imbalance exists for the hole populations left by photoexcitation along the Γ-X line [90,91]. The anisotropic electron and hole population can lead to an anisotropic screening of the valence potentials, deforming the lattice [92,93]. This has also been described as an anisotropic phonon bath deforming the lattice to distribute carriers between the degenerate but unequally occupied valleys [94–96]. The presence of the [100] lattice deformation could possibly correlate with the anharmonicity of the [100]-like optical phonon modes excited during *g*- and *f*-type inter-valley scattering.

In previous electron and x-ray diffraction measurements, the magnitude of the lattice deformation caused by the photoexcited carrier density has been approximated by an equilibrium stress ($\sigma_{DP,k}$) using [24,92]

$$\sigma_{DP,k} = -\sum_k \Delta N_k E_k \gamma_k = \sum_k \Delta N_k \frac{dE_k}{d\eta}, \qquad (11)$$

where $\Delta N_k$ is the number of excess carriers in a valley with momentum $k$, $E_k$ the energy of the band or electron at $k$, $\gamma_k = -\frac{1}{E_k}\frac{dE_k}{d\eta}$ is the Grüneisen coefficient, and $\frac{dE_k}{d\eta}$ is the electronic deformation potential for a valley with momentum $k$ and $\eta$ is the strain. For the thin silicon membranes used here, the strain in a particular $k$ direction after photoexcitation is approximated by

$$\eta_k = \frac{h_{100}}{B}\sigma_{DP,k}, \qquad (12)$$

where $h_{100} = 1 + \frac{2C_{12}}{C_{11}} = 1.77$ is a correction for the one-dimensionality of the strain [97] and $B$ is the bulk modulus of 97.6 GPa [98]. In Equation (11-12) a positive stress is taken as leading to a positive lattice expansion. From Reference [99], the deformation potential for a [100] strain is $\frac{dE_k}{d\eta} = 4.5\ eV$. A carrier density of 1-2x$10^{20}$ $cm^{-3}$ will therefore result in a lattice deformation of 0.1-0.3% expansion. According to the dipole selection rules for the indirect transition, the photoexcited carrier density is anisotropic along the [100] directions, and this would be the expected direction of the lattice deformation [86,87]. The predicted expansion range is intended to reflect the experimental uncertainty in the carrier density as well as the use of bulk quantities to represent the thin film membrane. While the estimated lattice deformation replicates the magnitude from the best fit of the data, Equation (12) is an equilibrium estimate to a very non-equilibrium effect. This approximation, coupled with the first order approximation of the electronic effects, means the exact origin of the lattice deformation in the experimental spectrum still warrants further exploration.

(***c***) ***2-4 Picoseconds*** An intermediate lineout at 2-4 ps is examined in the second to top panel of Figure 5b. At this intermediate time-scale, an isotropic lattice expansion could exist from the acoustic phonon modes excited by intra-valley carrier thermalization and the initial decay of phonons previously excited by inter-valley scattering. The carrier density is not yet reduced by Auger recombination, which occurs on a >10 ps time scale, however the carrier distribution should be equally distributed between degenerate valleys by electron-phonon scattering. Accordingly, the 2-4 ps lineout is found to be accurately predicted by the same carrier density as at 100 fs, 1.5±0.1 x $10^{20}$/$cm^3$, but by decreasing the lattice deformation to be 0.05±0.03% along the [100] direction (notation *100* in Figure 5b) and by adding a 0.06±0.01% isotropic expansion of all axes (notation of *111* in Figure 5b implies an equal, isotropic expansion of 0.06% along each axis). It should be noted that the absorption depth is large, ~10 microns for 800 nm excitation. This means that the carrier distribution is uniformly excited throughout the 200 nm membrane, and the subsequent stress and strain will also be uniform. The isotropic expansion dynamics therefore follow the frequency and anharmonicity of the excited phonons, as in a coherent phonon or non-thermal melting experiment. However, the fitting to both a volume expansion and deformation on this time scale, which is faster than the acoustic velocity, suggests a combined distortion without volume expansion [100].

(***b***) ***150 Picoseconds*** After the hot carrier and hot phonon baths are fully thermalized; a thermal isotropic lattice expansion should be the primary structural contribution to the differential absorption. The isotropic expansion results from the anharmonicity of acoustic phonon modes excited through intra-valley and phonon-phonon scattering, which is equivalent to heating the lattice [82]. Accordingly, the differential absorption at 150 ps (top panel in Figure 5b) is best fit by no [100] lattice deformation (0.00±0.08%), an isotropic expansion of 0.16±0.02%, and a reduced carrier density of 0.6±0.1 x$10^{20}$ carriers/$cm^3$ accounting for Auger recombination. A lattice thermalization accurately predicts the long time scale increase (decrease) in absorption at energies below (above) the dominant $L_1$ absorption peak. The increase in absorption around the $\Delta_1$ critical point feature as well as the appearance of an $L_3$ critical point feature with increasing time delay are also accurately predicted.

Following previous electron and x-ray diffraction measurements, the magnitude of the expected thermal lattice expansion can be estimated by summing over the effective Grüneisen parameters for the excited phonon modes in Equations (11-12) [24,92,101]. Using $\Delta N_k E_k = C_k \Delta T_k$ and summing over all involved phonon modes gives an isotropic stress of

$$\sigma_{Th} = -\sum_k \Delta N_k E_k \gamma_k = \gamma_v C_v(T)\Delta T = 3B\beta_v(T)\Delta T, \quad (13)$$

where $\beta_v(T)\sim 3\text{x}10^{-6}$ $K^{-1}$ is the linear thermal expansion coefficient [102], $B$ the bulk modulus, $\Delta T$ is the temperature change due to the thermalization of the photoexcited carrier density $\Delta N$, $C_v(T)$ is the heat capacity per unit volume of the involved phonon modes, and $\gamma_v$ is the sum over the relevant Grüneisen parameters [103,104]. The change in lattice temperature, $\Delta T$, is estimated by

$$\Delta T = \Delta N \frac{\hbar\omega}{C_v(T)}, \quad (14)$$

where $\Delta N$ is the excited carrier density, $\hbar\omega$ is the pump photon energy, $E_g$ is the band gap, and $C_v(T)$ the heat capacity per unit volume, here calculated within the Debye model using a Debye temperature of 645 K [103]. At the carrier densities used, VB-CB Auger recombination is also significant on this time scale. During Auger recombination already thermalized carriers are given additional energy by the three-body scattering process. This excess energy must be re-thermalized, leading to an additional heating of the lattice by 40-50 K on a 200 ps time-scale (see Section IV for further calculation details).

Including the heating from re-thermalized VB-CB Auger excited carriers, Equations (13-15) predict a 0.06-0.08% isotropic lattice expansion. The 0.1-0.2% thermal expansion range from the best fit is larger than the expansion predicted using Equation (13) by a factor of 2. The disagreement is most likely because of the equilibrium-approximation of Equation (13), but may also originate in the ultra-thin membrane geometry of the experiment, which has been reported to increase the actual thermal stress by 2-3 times [11,97,105]. It should also be noted that an isotropic lattice contraction is not observed as previously reported. The strength of the isotropic deformation is given by $\eta = d_{eh}\Delta N$, where $d_{eh}$=-1x10$^{-24}$ cm$^3$ [22,106]. For the experimental conditions here, this predicts an isotropic contraction that is 1/10$^{th}$ the strength of the predicted thermal expansion, explaining the absence of this effect.

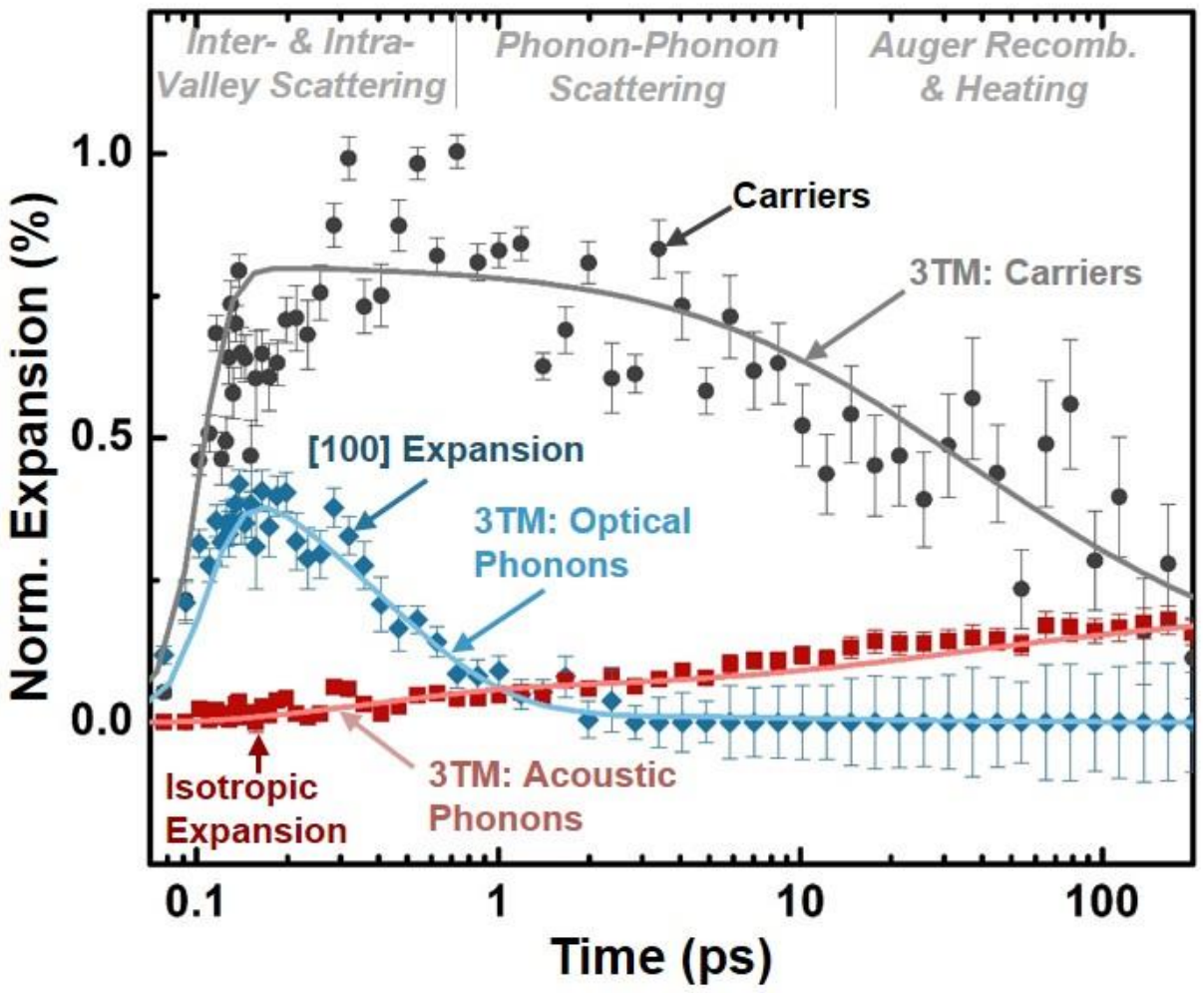

***Figure 6.*** *Time dependent carrier density, lattice deformation, and thermal lattice expansion as extracted from a non-linear fit of the measured excited state XUV absorption. The solid lines show the predictions of a three-temperature model (3TM) for the carrier density, optical phonons, and acoustic phonons. The points refer to the quantities extracted from the experimental data, with error bars representing the standard error of the fit to the experimental data. The non-linear fit is weighted by the experimental uncertainty. The scatter of the points is representative of the experimental noise. The lattice deformation is approximated as an*

*equilibrium [100] expansion and the lattice heating is approximated as an isotropic lattice expansion. The log scale of time is offset by 100 fs for visualization.*

## IV. QUANTIFYING CARRIER AND PHONON DYNAMICS

The results of Figure 5 suggest that the modeled electronic and structural effects describe the excited state XUV spectrum within the 0.2 eV experimental spectral resolution and 5 mOD experimental noise limit of the data at three key time scales. The fit procedure can equally be used to quantify the carrier density, lattice deformation, and lattice expansion for all times following 800 nm photoexcitation (Figure 6). The fit procedure should allow the kinetics of the photoexcited electrons, optical phonons, and acoustic phonons to be extracted accurately from the experimental XUV differential absorption. To validate that the fit kinetics match the previously measured kinetics of photoexcited silicon, the experimental fit values are compared to a three temperature model (3TM) in Figure 6 [8,10,107–112]. The solid lines show the 3TM predictions. The points refer to the experimentally extracted quantities, with error bars representing the standard error of the fit.

The 3TM is used to check the fit because it predicts the population averaged kinetics of the electrons, optical phonons, and acoustic phonons in photo-excited silicon, as well as the effects of thermal and carrier diffusion, Auger recombination, and the spatial dependence of the excited carrier distribution [8,10,107–112]. The 3TM is given as [108]

$$C_e \frac{\partial T_e}{\partial t} = \frac{D_a C_e}{n_e} \frac{\partial n_e}{\partial z} \frac{\partial T_e}{\partial z} - \frac{C_e}{\tau_{eo}} (T_e - T_o) - \frac{C_e}{\tau_{ea}} (T_e - T_a) + E_g \frac{n_e}{\tau_{Aug}(n_e)} + \left(\hbar\omega - E_g\right) * S(t, z), \tag{15}$$

$$C_o \frac{\partial T_o}{\partial t} = \frac{C_e}{\tau_{eo}} (T_e - T_o) - \frac{C_o}{\tau_{oa}} (T_o - T_a), \tag{16}$$

$$C_a \frac{\partial T_a}{\partial t} = \frac{\partial}{\partial z}\left(C_a D_t \frac{\partial T_e}{\partial z}\right) + \frac{C_e}{\tau_{ea}} (T_e - T_a) + \frac{C_o}{\tau_{oa}} (T_o - T_a), \tag{17}$$

$$\frac{\partial n_e}{\partial t} = \frac{\partial}{\partial z}\left(D_a \frac{\partial n_e}{\partial z}\right) - \frac{n_e}{\tau_{Aug}(n_e)} + S(t, z). \tag{18}$$

In Equations 16-19, $C_i$ and $T_i$ refer to the heat capacity per unit volume and temperature, respectively, of the electron ($e$), optical phonon ($o$), and acoustic phonon ($a$) baths. The temperatures of the electron and optical-phonon baths represent non-equilibrium temperatures of the excited state populations, not to be confused with the equilibrated lattice temperature. The electron-phonon and phonon-phonon coupling times between these baths are $\tau_{ea}$, $\tau_{eo}$, and $\tau_{oa}$. The electronic heat capacity is treated according to Reference [112], but it can be approximated as $\frac{3}{2} k_b n_e$ where $k_b$ is the Boltzmann constant. The temperature dependent heat capacity of the optical phonon modes is approximated by an Einstein model with an energy of 60 meV [113]. The temperature dependent heat capacity of the acoustic phonons is approximated using a Debye model with a Debye temperature of 645 K [114]. $D_a$ and $D_t$ refer to the ambipolar electronic and thermal diffusion constants, taken as 15 $cm^2/s$ and 0.88 $cm^2/s$ from References [115] and [116]. $D_a$ is approximated as a constant over the range of carrier densities present on the timescale of these experiments. $n_e$ is the spatially dependent carrier density created by the source term $S(t, z)$ which has a spatial and temporal profile matching the absorption depth of the excitation wavelength ($\hbar\omega$) in silicon ($E_g = 1.12\ eV$) and the pulse parameters reported in the Methods.

Using $\left(\hbar\omega - E_g\right) * S(t, x)$ as the source term of the 3TM implies that the kinetics are modeled only for thermalization to the band edge. If longer time scales are modeled up to when non-radiative recombination is present, Equation (15) should be $(\hbar\omega) * S(t, x)$ to imply that all photoexcited energy goes into heating the lattice and a non-radiative heating term should be included in Equation (17). Hot phonon bottleneck effects were not measured, so the optical and electronic temperatures are allowed to return to equilibrium independent of the acoustic phonon bath temperature. The Auger recombination time ($\tau_{Aug}$) is parameterized by the Richter model [117]. Impact ionization was tested for using the parameterization of Reference [112], but was not found to have a significant effect within the experimental noise limit of 5 mOD. Since the absorption depth and membrane thickness are smaller than the illuminated area, the spatial carrier and heating dynamics can be treated as one-dimensional [24]. Given that the thin silicon membrane is in vacuum and only radiative cooling is possible, the heat and carrier flux are treated as zero at both boundaries. The results from the 3TM are averaged over the XUV probe depth in the sample for comparison to experiment.

The 3TM prediction shown in Figure 6 uses an excitation density of $1.5\text{x}10^{20}$ $\text{cm}^{-3}$ and scattering times of $\tau_{eo}$, $\tau_{ea}$, and $\tau_{oa}$ equal to 30 fs, 500 fs, and 400 fs. These times are consistent with silicon's inter-valley scattering time of 20-60 fs [82,83], electron-acoustic phonon scattering time of 500 fs [107], and the 400 fs screened optical phonon lifetime at $10^{20}$ carriers/$\text{cm}^3$ excitation [118]. An inter-valley scattering time of 20-40 fs can reproduce the experimental rise time since the optical-phonon scattering time is within the excitation pulse width. The predicted carrier density from the 3TM rapidly rises within the pulse duration and then decays on a few ps time-scale due to Auger recombination, in agreement with the experimentally fit carrier density. For 800 nm excitation, the [100] wave-vector optical phonons take part in inter-valley and not intra-valley scattering [6,84,85]. Accordingly, the experimentally fit [100] lattice deformation mirrors the optical phonon kinetics predicted from the 3TM. The longer time scale isotropic lattice expansion kinetics are also accurately predicted by the lattice temperature from the 3TM.

The 3TM relates the average of the carrier and phonon scattering pathways between the initially excited carrier distribution and the final lattice temperature. The agreement between the fitted kinetics and the 3TM in Figure 6 therefore suggests that the XUV spectrum can be used to quantify the average inter-valley, intra-valley, and phonon-phonon scattering in relation to the carrier density. Of course, it must be remembered that the 3TM only represents the previously measured kinetics in silicon, and is an approximate treatment of the momentum-dependent non-equilibrium carrier and phonon populations present after excitation.

## VII. CONCLUSIONS

In conclusion, the transient XUV signal of the silicon 2p $L_{23}$ edge was analyzed in terms of possible electronic and structural changes following excitation in the Δ valley. The initially excited carrier distribution leads to state-filling, broadening, and band-gap renormalization. In the silicon XUV spectrum, the broadening induced changes are strongest, masking the state-filling, while the band-gap renormalization is offset by the core-hole exciton. On time scales up to a few picoseconds, an anisotropic lattice deformation dominates the structural contributions to the differential spectrum. This deformation initiates and decays on the timescale of excited optical phonon modes, matching the inter-valley thermalization kinetics of carriers following 800 nm excitation. On longer time scales, an isotropic lattice expansion corresponding to a lattice heating dominates the structural contributions to the differential XUV absorption. The kinetics of the lattice expansion match those of acoustic phonons involved in intra-valley thermalization and phonon-phonon scattering on a tens of picoseconds timescale. At >10 ps timescales, Auger recombination decreases the excited carrier density, decreasing the electronic contributions to the differential absorption but also further heating the lattice. By approximating these effects to first order using the BSE-DFT calculations of the ground state XUV absorption, the carrier and phonon dynamics following photoexcitation were quantified using a best-fit procedure. The fit kinetics agree well with previously measured photoexcited kinetics of silicon using a visible light probe. These experiments suggest that when the experimental XUV absorption can be accurately modeled in terms of the ground state properties of the material, transient XUV has potential for single-instrument determination of carrier and phonon scattering pathways in semiconductors.

## ACKNOWLEDGEMENTS

This work is supported by the Physical Chemistry of Inorganic Nanostructures Program, KC3103, Office of Basic Energy Sciences of the United States Department of Energy under Contract No. DE-AC02-05CH11231 through Materials Science and Engineering. SKC acknowledges support by the Department of Energy, Office of Energy Efficiency and Renewable Energy (EERE) Postdoctoral Research Award under the EERE Solar Energy Technologies Office. PMK acknowledges support from the Swiss National Science Foundation (P2EZP2_165252). MZ acknowledges support from the Humboldt Foundation.